\documentclass[a4paper, amsfonts, amssymb, amsmath, reprint, showkeys, showpacs, nofootinbib, twoside, superscriptaddress,aps,prl]{revtex4-1}
\usepackage[english]{babel}
\usepackage[utf8]{inputenc}
\usepackage{comment}
\usepackage[colorinlistoftodos, color=green!40, prependcaption]{todonotes}
\usepackage{appendix}
\usepackage{amsthm}
\usepackage{mathtools}
\usepackage{siunitx}
\usepackage{xcolor}
\usepackage{graphicx}
\usepackage{adjustbox}
\usepackage{placeins}
\usepackage[T1]{fontenc}
\usepackage{lipsum}
\usepackage{csquotes}
\usepackage{enumerate}
\usepackage{siunitx}
\usepackage{ulem}
\usepackage{hyperref}

\DeclareFontFamily{U}{mathb}{\hyphenchar\font45}
\DeclareFontShape{U}{mathb}{m}{n}{
      <5> <6> <7> <8> <9> <10> gen * mathb
      <10.95> mathb10 <12> <14.4> <17.28> <20.74> <24.88> mathb12
      }{}
\DeclareSymbolFont{mathb}{U}{mathb}{m}{n}
\DeclareMathSymbol{\Earth}{3}{mathb}{"43}

\begin{document}

\title{Bounds on screened dark energy from near-Earth space-based measurements}

\author{Fabiano Feleppa}
\email{ffeleppa@unisa.it}
\affiliation{Dipartimento di Fisica “E.R. Caianiello”, Università di Salerno, Via Giovanni Paolo II 132, I-84084 Fisciano, Italy}
\affiliation{INFN, Sezione di Napoli, Gruppo Collegato di Salerno, Italy}

\author{Welmoed Marit de Graaf}
\email{w.degraaf@studenti.unisa.it}
\affiliation{Dipartimento di Fisica “E.R. Caianiello”, Università di Salerno, Via Giovanni Paolo II 132, I-84084 Fisciano, Italy}

\author{Philippe Brax}
\email{philippe.brax@ipht.fr}
\affiliation{Institut de Physique Théorique, Université Paris-Saclay, CEA, CNRS, F-91191 Gif-sur-Yvette Cedex, France}

\author{Gaetano Lambiase}
\email{lambiase@sa.infn.it}
\affiliation{Dipartimento di Fisica “E.R. Caianiello”, Università di Salerno, Via Giovanni Paolo II 132, I-84084 Fisciano, Italy}
\affiliation{INFN, Sezione di Napoli, Gruppo Collegato di Salerno, Italy}


\begin{abstract}
   \noindent We test screened dark energy with near-Earth, space-based measurements. In a post-Newtonian framework, we compute leading corrections to geodetic precession (Gravity Probe B), LAGEOS-2 pericenter advance, and the Sagnac delay in a prospective orbital configuration, yielding bounds on chameleon, symmetron, and dilaton models. LAGEOS-2 sets the strongest Earth-orbit limits on symmetron and dilaton, while a Sagnac setup at the projected sensitivity of state-of-the-art space clocks gives the tightest chameleon constraint. These results show that low-density, space-based experiments sensitively probe screened dark energy and exclude previously allowed parameter space. Notably, at an assumed delay precision of $\mathcal{O}\big(10^{-19}\big)$, a Sagnac test would exclude the entire chameleon parameter space considered.
\end{abstract}


\maketitle

\noindent \textit{\textbf{Introduction. }}Understanding the physical origin of the late–time acceleration of the universe remains a central challenge in fundamental physics. The prevailing explanation — identifying the dark energy component of the universe with a cosmological constant — is empirically successful across multiple, independent probes (see, e.g., Refs.~\cite{Riess98,Perlmutter99,Nadathur2020,DiValentino2020,Escamilla2024}), yet it sharpens the naturalness problem associated with vacuum energy \cite{Carroll2001,Weinberg1989,Moreno-Pulido2022,Peracaula2022} and must contend with specific `tensions' within the standard $\Lambda$CDM (Lambda Cold Dark Matter) model \cite{Yang2018,Guo2019,Vagnozzi2020PRD,Visinelli2019,Alestas2020,DiValentino2021CQG,Dainotti2021,Vagnozzi2023Universe,DESI2025JCAP,CortesLiddle2024,Colgain2024,Carloni2025,GomezValent2024ApJ,Yang2024SciBull,Lodha2025DESI,Toda2024PDU,Giare2024JCAP,Jiang2024PRD,Giare2025PRD,GomezValent2025PLB,Giare2025PDU,CosmoVerse2025,RoyChoudhury2025,Scherer2025}. A well-motivated possibility is that the dark sector contains new light degrees of freedom and/or that gravity departs from General Relativity \cite{Ratra1988,Wetterich1988,Caldwell1998,Zlatev1999,Capozziello2002,Li2004,NojiriOdintsov2006,Bamba2012,ChamseddineMukhanov2013,Rinaldi2015,Dutta2018,Casalino2018,VisinelliVagnozzi2019,OdintsovOikonomou2019,Saridakis2020,OdintsovOikonomouPaul2020,Oikonomou2021,Motta2021,OikonomouGiannakoudi2022,Trivedi2024}; see also Ref.~\cite{Feoli2017} for a different perspective. However, if the new ingredient behind cosmic acceleration is a very light field that interacts with ordinary matter, it would generically produce an extra long-range “fifth force”, a signal that precision tests in the laboratory and in the Solar System have not detected. To remain consistent with those tests, the new interaction must be effectively hidden locally while still influencing cosmic dynamics on large scales. This is the idea of screening: in dense environments the extra force is suppressed, whereas on cosmological scales it can remain active (for details, see the recent review \cite{Brax2021Universe}). Very well-studied examples include the chameleon mechanism~\cite{KhouryWeltman2004PRL,KhouryWeltman2004PRD,Brax2004a,Brax2004b,Capozziello2008,Brax2008,Brax2010a,Brax2010b,Gannouji2010,Brax2012,Wang2012,Khoury2013,Erickcek2014,Elder2016,Brax2016,Burrage2016,Burrage20182,Borràs2019,Burrage2019,Sakstein2019,Desmond2019,Uzan2020,Vagnozzi2020MNRAS,Cai2021,Vagnozzi2021,Borràs2021,Karwal2022,Katsuragawa2022,Dima2021,Benisty2022,Tamosiunas2022a,Briddon2021,Brax20221,Ferlito2022,Yuan2022,Tamosiunas2022b,Brax20231,Boumechta2023,Benisty2023b,Paliathanasis2023a,Paliathanasis2023b,Benisty2023c,Briddon2024,Hogas2023,Benisty2024AandA,Kumar2024,BaezCamargo2024,Lévy2024,OShea2024,Hartley2024,Elder2024,Paliathanasis2024,Pizzuti2024,Yuan2025} (the field becomes short-ranged as the surrounding density rises), the symmetron~\cite{HinterbichlerKhoury2010,Khoury2011,Davis2012ApJ,Brax2011PRD,Winther2012ApJ,Upadhye2013PRL,Bamba2013JCAP,Silva2014PRD,Burrage2016JCAP,Burrage2017PRD,Pinho2017PLB,OHare2018PRD,Burrage2019PRD,Contigiani2019PRD,Cronenberg2018NatPhys,Elder2020PRD,Jenke2021EPJST,Pitschmann2021PRD,Perivolaropoulos2022PRD,Nosrati2023PRD,Christiansen2023JCAP,Hogas2023PRD,Kading2023Astronomy,Christiansen2024AandA,Xiong2025PRD,Li2025JCAP} and dilaton~\cite{DamourPolyakov1994,Brax20102,Brax2011,Brax2012JCAP,Brax2017PRD,Hartley2019PRD,Burgess2022JCAP,Brax2023JCAP,Fischer2024PTEP,FischerSedmik2024JCAP,Reyes2024PRD,Fischer2025PDU,Smith2024JCAP,Smith2024arXiv,Kading2025PDU,Smith2025arXiv,BachsEsteban2025Universe,Brax2025W} mechanisms (the coupling to matter switches off in high-density regions), the K-mouflage mechanism \cite{Babichev2009,Valageas2014} and the Vainshtein effect \cite{Vainshtein1972} (nonlinear derivative self-interactions amplify the effective kinetic term and suppress scalar gradients around massive bodies, screening the force within a characteristic K-mouflage/Vainshtein radius). These frameworks have been developed extensively and tested against a broad range of measurements — from torsion-balance and atom-interferometry experiments to lunar and satellite laser ranging, and other astrophysical and cosmological observations — yielding tight constraints while still leaving viable regions of parameter space to explore \cite{Burrage2018}.

Screened modified gravity can be cast in the language of scalar–tensor theories \cite{Sakstein2014,Zhang2016}. In the case of weak gravitational fields, one can consider the Parameterized Post-Newtonian (PPN) expansion, characterized by an effective gravitational strength $\mathcal{G}$ and the PPN coefficients $\gamma$ and $\beta$.~Screening shows up as an environment/source dependence of these quantities \cite{Zhang2016} — high-density regions drive $\mathcal{G} \to G$, $\gamma \to 1$, and $\beta \to 1$, keeping the extra force hidden, while lower-density can allow for small departures. Unlike purely laboratory tests, space-based measurements are crucial because the ambient density in near-Earth is many orders of magnitude lower, which can reduce screening and enhance sensitivity to deviations.

In the context of high–precision space–based tests of General Relativity, landmark experiments include Gravity Probe B \cite{Buchman2000,Everitt2015}, Lunar Laser Ranging \cite{Bertotti1987,Williams1996}, LARES-2 \cite{Capozziello2015,Ciufolini2023}, LAGEOS-2 \cite{Iorio2004,Ciufolini2004,Iorio2009}, very-long-baseline interferometry (VLBI) \cite{Schuh2012,Otsubo2020}, the GINGER ring-laser program \cite{Altucci2023,DiVirgilio2024}, and the GRACE/GRACE-FO gravity missions \cite{Tapley2004,Abich2019}, among others. These experiments tighten bounds on the PPN coefficients $\gamma$ and $\beta$ and on associated non-Keplerian effects, thereby constraining broad classes of fifth-force and modified-gravity scenarios (see, e.g., Refs.~\cite{Lambiase2013,Radicella2014,Mariani2024,Lévy20242}), including screened models; with current timing capabilities, Sagnac-type orbital configurations using space-qualified clocks also emerge as sensitive probes of tiny departures from General Relativity in Earth’s field, complementing the tests above \cite{Aliberti2025}.~In the following, we investigate screened dark energy using a selection of the space–based tests outlined above; within the formalism of Ref.~\cite{Zhang2016} at first post-Newtonian order (1PN) (see also Ref.~\cite{Feleppa2024}), we obtain the leading corrections induced by chameleon, symmetron, and dilaton to three key observables:~(i) the geodetic precession of a gyroscope as measured by Gravity Probe B, (ii) the pericenter precession inferred from LAGEOS-2, and (iii) the Sagnac time delay in a prospective setup employing state-of-the-art space clocks; we then map these predictions onto the screening parameter space and compare them with Solar-System and laboratory bounds. This yields new constraints, thereby excluding previously allowed regions of parameter space. In what follows, we work in units with $c=\hbar=1$ and with signature convention $\{-,+,+,+\}$.
\newline
\noindent \textbf{\textit{Screened scalar-tensor theories. }}At 1PN order, the Jordan-frame line element for a static, spherically symmetric source of mass $M$ can be written as \cite{Will2014}
\begin{equation}
    \begin{aligned}
    ds^2 &= - \left ( 1 - \frac{2\,\mathcal{G} M}{r}\,+\,\beta\frac{2\,\mathcal{G}^2 M^2}{r^2} \right ) dt^2 \\
    &\quad \quad \quad + \left ( 1\,+\,\gamma \frac{2\,\mathcal{G} M}{r} \right ) \left ( dr^2 + r^2d\Omega^2 \right ) \,,
    \end{aligned}
\label{eq:ds2pot}
\end{equation}
where $d\Omega^2 \equiv d\vartheta^2 + \sin^2\vartheta\, d\varphi^2$ is the metric on the unit two-sphere, $\mathcal{G}$ is the effective gravitational strength, and finally $\gamma$ and $\beta$ are the PPN parameters. In screened models, these quantities depend on the ambient background density $\rho_\infty$ (here taken as the local Galactic density) through $\phi_\infty$, the scalar field value far from the source. In what follows, we consider three representative screening mechanisms:~chameleon, symmetron, and dilaton. For the chameleon mechanism, the effective gravitational strength $\mathcal{G}$, $\gamma$, and $\beta$ are given by \cite{Zhang2016,Feleppa2024}
\begin{align}
    \mathcal{G}_{\text{cham}} &= G \left ( 1 + \frac{\beta_m \phi_{\infty}}{M_{\text{Pl}}\Phi_{\text{N}}} \right ) \,, \label{eq:Gepscham} \\
    \gamma_{\text{cham}} &= 1 - \frac{2\beta_m \phi_{\infty}}{M_{\text{Pl}}\Phi_{\text{N}}}\,,
    \label{eq:gammaepscham} \\
    \beta_{\text{cham}} &= 1 - \frac{3}{4\Phi_{\text{N}}(n + 1)} \left ( \frac{\phi_{\infty}}{M_{\text{Pl}}} \right ) ^2\,,
\label{eq:betaepscham}
\end{align}
respectively, where $\Phi_{\text{N}} = GM/R$ is the Newtonian potential, with $R$ being the gravitating source's radius, and
\begin{equation}
    \phi_\infty = \left ( \frac{n M_{\text{Pl}}\Lambda^{4 + n}}{\beta_m \rho_\infty} \right ) ^{\frac{1}{n + 1}}\,,
\label{eq:phiinftycham}
\end{equation}
where $\Lambda$ sets the mass scale, $\beta_m > 0$ is the dimensionless matter coupling strength of the scalar, and $n > 0$.

For the symmetron, $\mathcal{G}$, $\gamma$, $\beta$, and $\phi_{\infty}$ are \cite{Zhang2016,Feleppa2024}
\begin{align}
    \mathcal{G}_{\text{sym}} &= G \left ( 1 + \frac{\phi_{\infty}^2}{M_{\text{sym}}^2\Phi_{\text{N}}} \right ) \,,
    \label{eq:Gepssym} \\
    \gamma_{\text{sym}} &= 1 -     \frac{2\phi_{\infty}^2}{M_{\text{sym}}^2\Phi_{\text{N}}}\,,
    \label{eq:gammaepssym} \\
    \beta_{\text{sym}} &= 1 + \frac{1}{2} \left ( \frac{\phi_{\infty}}{M_{\text{sym}}\Phi_{\text{N}}} \right ) ^2\,,
    \label{eq:betaepssym}
    \\
    \phi_{\infty} &= \frac{\mu}{\sqrt{\lambda}} \sqrt{1 - \frac{\rho_{\infty}}{M_{\text{sym}}^2\mu^2}}\,,
    \label{eq:phiinftysymmetron}
\end{align}
where $\mu$ and $M_{\text{sym}}$ are mass-dimension one parameters, while $\lambda$ is a positive, dimensionless coupling constant. The above results hold in the regime $\rho_{\infty} < M_{\text{sym}}^2\mu^2$.

Finally, the dilaton case is characterized by \cite{Zhang2016,Feleppa2024}
\begin{align}
    \mathcal{G}_{\text{dil}} &= G \left ( 1 + \frac{A_2 \, \phi_\infty^2}{M_{\text{Pl}}^2\Phi_{\text{N}}} \right ) \,,
    \label{eq:Gepsdil} \\
    \gamma_{\text{dil}} &= 1 - \frac{2 A_2 \, \phi_\infty^2}{M_{\text{Pl}}^2\Phi_{\text{N}}}\,,
    \label{eq:gammaepsdil} \\
    \beta_{\text{dil}} &= 1 + \frac{A_2}{2} \left ( \frac{\phi_\infty}{M_{\text{Pl}}\Phi_{\text{N}}} \right ) ^2\,,
    \label{eq:betaepsdil} \\
    \phi_\infty &= \frac{\lambda_{\text{dil}} \, M_{\text{Pl}} V_0 }{A_2 \rho_\infty} \,,
    \label{eq:phimindil}
\end{align}
where $V_0$ is a constant with units of energy density, $\lambda_{\text{dil}}$ is the self-coupling, and $A_2$ is the parameter controlling the strength of the coupling of the scalar to matter \cite{Voith2023}.

It is worth noting the extreme similarity between the PPN parameters of the symmetron and dilaton where $M_{\rm sym} \equiv M_{\rm Pl}/\sqrt{A_2}$. The difference between the two models arises from their dependence on $\rho_\infty$ via $\phi_\infty$, and the resulting difference can be significant.

Further details on the framework, as well as of the approximations made to derive the equations discussed here are provided in the Supplemental Material (SM).

\noindent \textbf{\textit{Solar-System tests. }}Here we present formulas for the three observables used to constrain screened modified gravity:~the geodetic precession of a gyroscope (as measured by Gravity Probe B \cite{Buchman2000,Everitt2015}), the secular pericenter–precession rate of LAGEOS-2 \cite{Iorio2004,Ciufolini2004,Iorio2009}, and the Sagnac time delay for an idealized experiment around a spherically symmetric central mass (the Earth, in our case) \cite{Aliberti2025}. We also present our constraint methodology. For detailed descriptions of these three experiments and the derivations, see the SM.

\noindent \textit{Geodetic precession. }Starting from the metric \eqref{eq:ds2pot}, the geodetic precession per orbit at 1PN is found to be
\begin{equation}
     \alpha_{\text{screen}} = \pi \left( 1 + 2\gamma \right)\frac{\mathcal{G} M}{R_{\text{orb}}}\,,
\label{eq:precessionscreen}
\end{equation}
where we recall that the effective gravitational strength $\mathcal{G} \to (\mathcal{G}_{\text{cham}},\mathcal{G}_{\text{sym}},\mathcal{G}_{\text{dil}})$ and $\gamma \to (\gamma_{\text{cham}},\gamma_{\text{sym}},\gamma_{\text{dil}})$, depending on the mechanism under consideration; by multiplying $\alpha$ by the number of orbits per year, we obtain the drift rate measured by Gravity Probe B (GP-B), which has an associated relative uncertainty of approximately $0.28\%$ \cite{Everitt2011}. Any correction to the rate due to the presence of a screened scalar field must remain within this limit to ensure consistency; hence we require
\begin{equation}
    \left | \frac{\alpha_{\text{screen}} - \alpha_{\text{GR}}}{\alpha_{\text{GR}}} \right | \lesssim 2.8 \times 10^{-3},
\label{eq:deviationGPB}
\end{equation}
where $\alpha_{\text{GR}} \equiv \alpha_{\text{screen}}\rvert_{\gamma \to 1}$.

We note that Earth’s oblateness affects GP-B both through gravity-gradient torques (modeled \cite{Everitt2011}) and through a relativistic quadrupole correction to the geodetic drift; the latter is $\sim 10^{-3}$ of the monopole term \cite{Adler2000}. Since Earth’s $J_2$ is known at the $\sim 10^{-7}$ level \cite{Paris2024}, the induced fractional uncertainty turns out to be $\sim 10^{-10}$, negligible compared to the GP-B uncertainty used.

\noindent \textit{Pericenter advance. }At 1PN order, the secular pericenter–precession rate for a test body orbiting a spherically symmetric central mass in scalar–tensor gravity is \cite{Will2014}
\begin{equation}
    \Delta \dot{\omega}^{\text{screen}} = \frac{3\left(\mathcal{G} M\right)^{3/2}}{a^{5/2}(1-e^2)}\frac{2+2\gamma - \beta}{3}\,,
\label{eq:secularratescreen}
\end{equation}
where $a$ denotes the semi-major axis of the ellipse and $e$ is the eccentricity of the orbit. Furthermore, we recall that screening effects are encoded in the gravitational strength $\mathcal{G}$, $\gamma$, and $\beta$:~$\mathcal{G} \to (\mathcal{G}_{\text{cham}},\mathcal{G}_{\text{sym}},\mathcal{G}_{\text{dil}})$, $\gamma \to (\gamma_{\text{cham}},\gamma_{\text{sym}},\gamma_{\text{dil}})$, and $\beta \to (\beta_{\text{cham}},\beta_{\text{sym}},\beta_{\text{dil}})$, depending on the specific mechanism considered. We constrain the screening parameter space by requiring that the secular pericenter-precession rate predicted by the screened model does not differ from the General Relativity prediction by more than the relative uncertainty of the LAGEOS-2 measurement \cite{Lucchesi2010}:
\begin{equation}
    \left | \frac{\Delta \dot{\omega}^{\text{screen}} - \Delta \dot{\omega}^{\text{GR}}}{\Delta \dot{\omega}^{\text{GR}}} \right | \lesssim 2.1 \times 10^{-3}\,,
\label{eq:deviationLAGEOS}
\end{equation}
where $\Delta\dot{\omega}_{\text{GR}} \equiv \Delta\dot{\omega}_{\text{screen}}\rvert_{\gamma \to 1,\,\beta \to 1}$.

It is important to point out that, in orbit determinations, the product $G M$ is solved directly from the same tracking data:~any orbit-constant change in the gravitational strength due to screening is simply absorbed into that fitted parameter and cancels in the ratio of secular rates. Consequently, the LAGEOS-2 observable is insensitive to $\mathcal G$ itself and constrains only the PPN factor $\big(2+2\gamma-\beta\big)/3$. This modeling proviso also applies to GP-B and Sagnac, which thus probe $\gamma$ only.

Moreover, we also point out that, in the LAGEOS-2 analysis, the Earth's quadrupole $J_2$ is explicitly modeled and subtracted \cite{Lucchesi2010}.

\noindent \textit{Sagnac time delay. }The Sagnac experiment proposed in Ref.~\cite{Aliberti2025} is conceived as a space-based analog of a ring laser gyroscope, but with the closed path given by a satellite’s orbit around a central body (the Earth, in our case). A satellite in circular orbit is imagined to carry a precision atomic or nuclear clock, which acts as the reference against which light travel times can be measured. Two light beams are sent simultaneously in opposite directions along the orbital loop:~one co-rotating with the orbital motion, the other counter-rotating. After completing a full revolution, the beams return to the satellite, where the onboard clock records their arrival times. Starting from the 1PN metric in Eq.~\eqref{eq:ds2pot}, the Sagnac time difference between the two signals, expressed in the satellite’s proper time $\tau$, is
\begin{equation}
     \Delta \tau_{\text{screen}} = 4\pi R_{\text{orb}}^2 \Omega\left[ 1 + \left( 1 + 2\gamma \right)\frac{\mathcal{G} M}{R_{\text{orb}}}{}+v^2/2 \right]\,,
\label{eq:timedifferencescreen}
\end{equation}
where $\Omega$ is the satellite’s constant angular velocity and $R_{\text{orb}}$ its orbital radius. The above formula is valid in the slow-rotation regime, $v \equiv R_{\text{orb}}\,\Omega \ll 1$.

Proceeding as in the two previous cases, the screening parameter space can be constrained by requiring the relative deviation from the General Relativity prediction in the Sagnac time shift to be less than the experimental relative uncertainty. As previously mentioned, since no orbital Sagnac experiment of the type proposed has yet been performed, there are no direct data available to determine its achievable precision. In order to set indicative bounds, one can instead rely on the performance of present-day space-qualified atomic clocks, which provide the natural reference for measuring timing differences in such a setup. Current devices such as the Deep Space Atomic Clock (DSAC) \cite{Delva2018} and the PHARAO clock aboard the ACES mission \cite{Burt2021} reach fractional frequency uncertainties in the range of $10^{-15}-10^{-16}$ over relevant integration times (below, we choose $10^{-15}$ for a conservative estimate). As an idealized benchmark, assuming that the Sagnac delay can be measured with the same fractional precision, we adopt this level of sensitivity as the relative uncertainty for our prospective measurement. Therefore, we can write
\begin{equation}
    \left | \frac{\Delta \tau_{\text{screen}} - \Delta \tau_{\text{GR}}}{\Delta \tau_{\text{GR}}} \right | \lesssim 10^{-15} \, ,
\label{eq:deviationSagnac}
\end{equation}
where $\Delta\tau_{\text{GR}} \equiv \Delta\tau_{\text{screen}}\rvert_{\gamma \to 1}$.

Above, we did not include the effect of the quadrupole moment $J_2$. In the SM we estimate the $J_2$ error budget for the Sagnac case; since the fractional impact of the $J_2$ uncertainty on the Sagnac time delay (normalized to the General Relativity baseline) turns out to be below $10^{-15}$, it does not affect our bounds. In other words, if any modeling residual from $J_2$ exceeded $10^{-15}$, the right-hand side of Eq.~\eqref{eq:deviationSagnac} should be replaced by the total residual error budget, which would accordingly weaken the bound. In SM, we show that this is not the case.

\noindent \textbf{\textit{Results. }}Let us specialize Eqs.~\eqref{eq:deviationGPB}, \eqref{eq:deviationLAGEOS} and \eqref{eq:deviationSagnac} to the three screening mechanisms of interest. For experiments around the Earth, we have $\Phi_N = \Phi_{N,\Earth} \sim 7.0 \times 10^{-10}$; moreover, we approximate the local Galactic ambient density as $\rho_{\infty} \sim 10^{-6} \,\text{eV}^4$ \cite{Zhang2016,KhouryWeltman2004PRL}.

\noindent \textit{Chameleon screening. }Here we adopt the widely used dark-energy normalization $\Lambda = \Lambda_{\text{DE}} \sim 2.4\,\text{meV}$ \cite{Feleppa2024}. This choice ensures that the chameleon field can act like a quintessence component and generate cosmic acceleration on large scales. It simultaneously simplifies the parameter space so that only the power index $n$ and the coupling $\beta_m$ need to be specified. Existing bounds on $n$ and $\beta_m$ can be found, e.g., in Fig.~6 of Ref.~\cite{Burrage2018}. As shown there, for $n \geq 1$ a large part of the parameter space is already excluded by torsion balance measurements \cite{Upadhye2012}, neutron interferometry \cite{Lemmel2015,Li2016}, and astrophysical constraints (Cepheid variable stars \cite{Jain2013} and rotation
curve tests \cite{Vikram2018}), leaving the window $1 \lesssim \beta_m \lesssim 100$ still viable. Motivated by this, we concentrate on the relatively unconstrained region given by $0.1 \lesssim n \lesssim 1$ and $1 \lesssim \beta_m \lesssim 100$. A plot similar to Fig.~6 of Ref.~\cite{Burrage2018} can be found in Fig.~1(a) of Ref.~\cite{Fischer2024}, which also illustrates the improvement in interferometric bounds obtained by incorporating the projected sensitivity of CANNEX.
\begin{figure}[!t]
\includegraphics[width=0.7\linewidth]{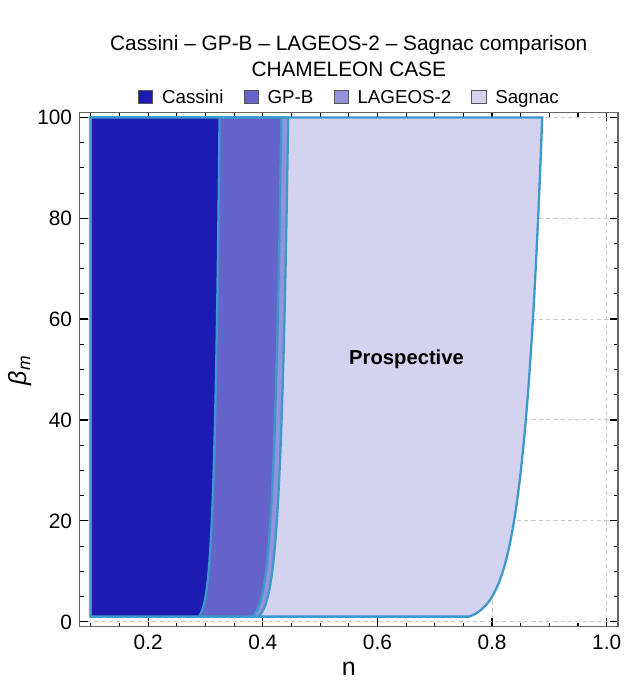}
\centering
\caption{Constraints on the chameleon parameter space in the plane $(n, \beta_m)$ for $\Lambda = \Lambda_{\text{DE}} \sim 2.4\,\text{meV}$. The shaded (colored) regions indicate the portions of parameter space that are excluded by the different experiments considered here, together with the Solar System light-deflection constraint derived from the Cassini experiment \cite{Zhang2016}. In the Sagnac case, we assume a satellite orbit radius $R_{\text{orb}}$ corresponding to an altitude of approximately $500\,\text{km}$ above Earth’s surface.}
\label{fig:comparisonchameleon}
\end{figure}

The constraints in our case can be inferred by plugging Eqs.~\eqref{eq:Gepscham}, \eqref{eq:gammaepscham}, and \eqref{eq:betaepscham} into Eqs.~\eqref{eq:deviationGPB}, \eqref{eq:deviationLAGEOS} and \eqref{eq:deviationSagnac}.
In Fig.~\ref{fig:comparisonchameleon}, we show a contour plot comparing the constraints from the three experiments considered here, together with the constraint from Cassini (laboratory bounds are not shown because they do not exclude any portion of the parameter space presented here \cite{Burrage2018}).

It is worth commenting on how the strength of the constraints depends on the source potential of the experiment. It is established that Solar System light-deflection experiments such as Cassini provide the tightest existing limits on the PPN parameter $\gamma$. However, these measurements probe the gravitational field of the Sun, whose deep potential well renders it strongly self-screened. Because the Sun is much more compact than the Earth, the two bodies exhibit very different screening efficiencies:~the Earth’s surface potential is about three orders of magnitude smaller, making it far less effective at suppressing scalar charges. As a consequence, the effective $\gamma$ parameter probed by near-Earth experiments such as GP-B can deviate much more strongly from unity than the value probed in the solar field. Thus, although the Cassini constraint on $\gamma$ is tighter, its translation into bounds on model parameters can make Earth-orbit experiments equally or more constraining; this contrast in source potentials highlights why GP-B constitutes a powerful probe of screening mechanisms.~A similar reasoning applies to the LAGEOS-2 experiment. As for the Sagnac experiment, which would provide the strongest constraint (see Fig.~\ref{fig:comparisonchameleon}), this advantage arises from the high precision with which time differences can be measured, thanks to rapid advances in atomic-clock technology.

\noindent \textit{Symmetron screening. }Here we examine the subspace spanned by $M_{\text{sym}}$ and $\lambda$, fixing the tachyonic mass $\mu$ to two representative values:~\mbox{$\mu = 10^{-22}\,\text{eV}$} and \mbox{$10^{-19}\,\text{eV}$}. Specifically, we focus on the ranges \mbox{$-76 \le \log_{10}\lambda \le -52$} and \mbox{$10.5 \le \log_{10}(M_{\text{sym}}/\text{GeV}) \le 11.5$}. As inferred from Fig.~10 of Ref.~\cite{Burrage2018}, Fig.~2 of Ref.~\cite{Fischer2024}, and from the constraint from Mercury’s perihelion shift~\cite{Zhang2016}, this region remains largely unconstrained. Indeed, torsion-balance experiments~\cite{Upadhye2013PRL}, atom interferometry~\cite{Jaffe2013}, precision neutron tests~\cite{Fischer2024}, spectroscopy of hydrogen and muonium, the electron (g-2)~\cite{Brax20231}, gravity-resonance spectroscopy (qBounce)~\cite{Pitschmann2018,Cronenberg2018NatPhys,Pitschmann2021PRD,Fischer2024}, and the projected sensitivity of CANNEX~\cite{Fischer2024} probe much larger values of~$\mu$.
\begin{figure*}[!t]
\centering
\includegraphics[width=0.324\textwidth]{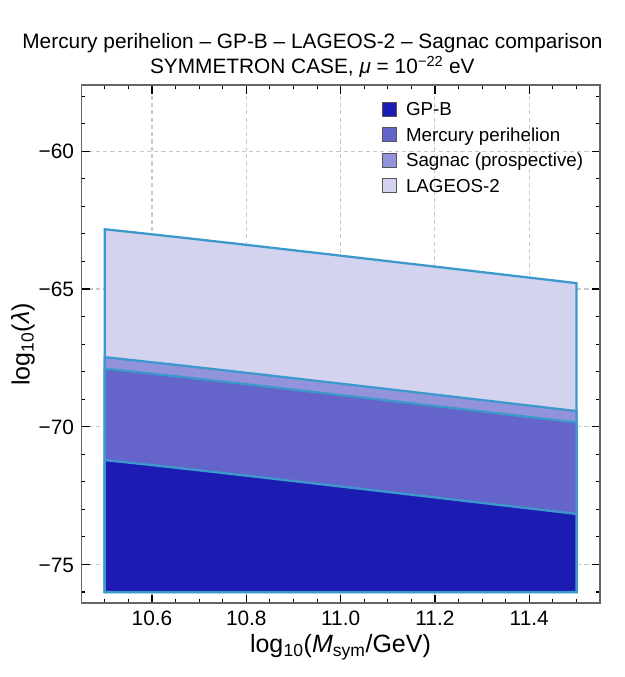}\hspace{2cm}\includegraphics[width=0.324\textwidth]{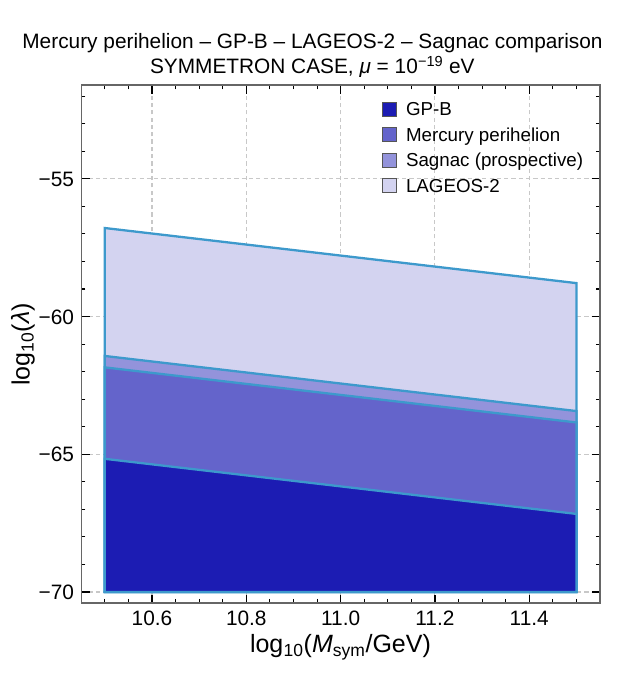}
\caption{Constraints on the symmetron parameter space in the plane $(\log_{10}(M_{\text{sym}}/\mathrm{GeV}),\log_{10}(\lambda))$ with $\mu=10^{-22}\,\mathrm{eV}$ (left) and $10^{-19}\,\mathrm{eV}$ (right). Both axes are shown on a base-10 logarithmic scale. Shaded (colored) regions indicate exclusions from GP-B, LAGEOS-2, Sagnac, and the bound from Mercury’s perihelion shift~\cite{Zhang2016}. For Sagnac, we assume $R_{\mathrm{orb}} \sim R_\Earth + 500\,\mathrm{km}$.}
\label{fig:comparisonsymmetron}
\end{figure*}

\begin{figure*}[!t]
\centering
\includegraphics[width=0.324\textwidth]{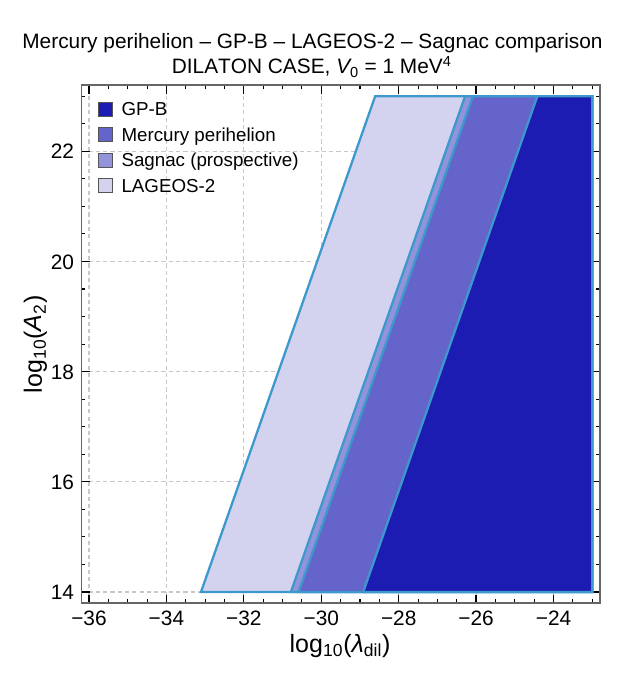}\hspace{2cm}\includegraphics[width=0.324\textwidth]{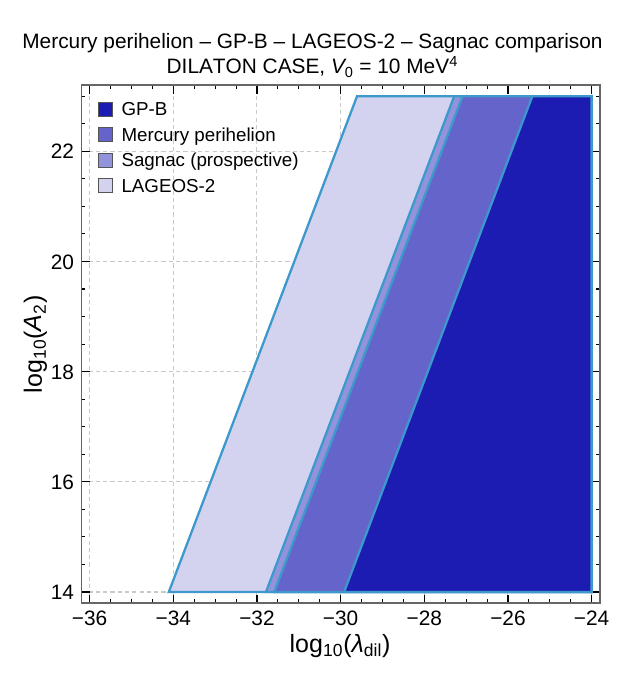}
\caption{Constraints on the dilaton parameter space in the plane $(\log_{10}(\lambda_{\text{dil}}),\log_{10}(A_2))$ with \mbox{$V_0=1\,\mathrm{MeV^4}$} (left) and \mbox{$V_0=10\,\mathrm{MeV^4}$} (right). Both axes are shown on a base-10 logarithmic scale. Shaded (colored) regions indicate exclusions from GP-B, LAGEOS-2, Sagnac, and the bound from Mercury’s perihelion shift~\cite{Zhang2016}. For Sagnac, we assume $R_{\mathrm{orb}} \sim R_\Earth + 500\,\mathrm{km}$.}
\label{fig:comparisondilaton}
\end{figure*}

Constraints on the parameter space are obtained by substituting Eqs.~\eqref{eq:Gepssym}, \eqref{eq:gammaepssym}, and \eqref{eq:betaepssym} into Eqs.~\eqref{eq:deviationGPB}, \eqref{eq:deviationLAGEOS}, and \eqref{eq:deviationSagnac}. In Fig.~\ref{fig:comparisonsymmetron}, we show two contour plots (fixing $\mu=10^{-22}\,\text{eV}$ and $\mu=10^{-19}\,\text{eV}$) comparing the constraints from the three experiments considered, alongside the bound from Mercury’s perihelion shift~\cite{Zhang2016} (also in this case, we omit laboratory limits, as they do not restrict the parameter region displayed \cite{Burrage2018,Fischer2024}). As $\mu$ increases, the model moves deeper into the symmetry-broken regime and the ambient field value rises. Screening becomes less efficient, so the predicted deviations grow; accordingly, the exclusion contours shift upward and to the right (toward larger $M_{\text{sym}}$ and $\lambda$) and the allowed region shrinks. That LAGEOS-2 yields the tighter bound is expected in symmetron gravity:~deviations in $\gamma$ are suppressed by an extra factor of the source’s compactness $\Phi_N$ relative to those in $\beta$ (see Eqs.~(78) and (79) of Ref.~\cite{Zhang2016}). As a result, tests that constrain $\beta$ via pericenter advance dominate the parameter bounds, whereas $\gamma$-only probes (e.g., light deflection/Shapiro delay as in Cassini) are intrinsically less powerful in this class of models. Consequently, although Earth’s smaller $\Phi_N$ partially enhances the effective symmetron-induced shift in $\gamma$ probed by GP-B and Sagnac, this boost is insufficient to overcome the superior leverage of a pericenter-advance measurement.~By contrast, LAGEOS-2 tests a $\beta$-sensitive observable (Earth-orbit pericenter precession), and is thus more effective than GP-B within the same slice of parameter space.

\noindent \textit{Dilaton screening. }Considering bounds from Mercury’s perihelion shift \cite{Zhang2016}, precision neutron tests \cite{Fischer2024PTEP}, Lunar Laser Ranging \cite{Sedmik2024}, qBounce and CANNEX \cite{Fischer2024} (the last of which is prospective), we focus here on the largely unconstrained region $-36 \le \log_{10}\!\lambda_{\rm dil} \le -23$ and $14 \le \log_{10}\!A_2 \le 23$. In what follows, we fix $V_0$ at two representative values, $V_0 = 1\,\mathrm{MeV}^4$ and $V_0 = 10\,\mathrm{MeV}^4$; for the cosmological motivation of this model, see Ref.~\cite{Brax2022}, where $V_0$ is tied to the dark-energy density.

As in the previous cases, constraints on the dilaton parameter space can be found by inserting Eqs.~\eqref{eq:Gepsdil}, \eqref{eq:gammaepsdil} and \eqref{eq:betaepsdil} into Eqs.~\eqref{eq:deviationGPB}, \eqref{eq:deviationLAGEOS}, and  \eqref{eq:deviationSagnac}. Fig.~\ref{fig:comparisondilaton} shows the region of dilaton parameter space of interest. The two panels correspond to two different values of $V_0$, as discussed above; increasing $V_0$ raises the background minimum $\phi_\infty$ linearly (see Eq.~\eqref{eq:phimindil}), weakening screening effects. Consequently, the allowed region shrinks when going from $V_0 = 1\,\mathrm{MeV}^4$ to $V_0 = 10\,\mathrm{MeV}^4$. As in the symmetron case, we expect the LAGEOS-2 bounds to be the most stringent:~pericenter advance probes the $\beta$-combination with a leading correction that scales as $1/\Phi_N^2$, whereas GP-B and Sagnac are essentially $\gamma$–tests with a $1/\Phi_N$ scaling; hence, the LAGEOS-2 constraint dominates (see Eqs.~(92)–(93) in Ref.~\cite{Zhang2016}).

\noindent \textbf{\textit{Conclusions. }}In the symmetron and dilaton cases, the current leader among the tests is LAGEOS-2; this follows from the observable’s $\beta$-sensitivity and its stronger scaling with $\Phi_{\text{N}}$.~For the chameleon case, the projected Sagnac measurement provides the tightest constraint among the tests we considered.~Crucially, if the experimental sensitivity on the delay reaches the $\mathcal{O}(10^{-19})$ level of nuclear clocks~\cite{Campbell2012}, the Sagnac experiment becomes decisively powerful:~within our modeling assumptions, such a measurement would exclude the entire region of chameleon parameter space we considered. Our results underline that experiments in low Earth orbit provide sensitive probes of screened modified gravity. 

\noindent \textbf{\textit{Acknowledgments. }}We are grateful to Jo\"el Berg\'e and Clare Burrage for valuable comments on the manuscript. FF and GL acknowledge support from the University of Salerno and the Istituto Nazionale di Fisica Nucleare (INFN) through the Commissione Scientifica Nazionale 4 (CSN4) Iniziativa Specifica ``Quantum Universe''.

\section{Supplemental Material}

\noindent\textbf{\textit{Screened scalar-tensor theories.~}}Closely following Refs.~\cite{Zhang2016,Feleppa2024}, here we review the scalar–tensor framework used in the main text to evaluate screened-gravity corrections to the three observables of interest.

Scalar–tensor theories can be formulated in two frames. In the Jordan frame (JF), matter couples minimally to $\tilde g_{\mu\nu}$ and the action has Brans–Dicke form \cite{Faraoni1999}
\begin{multline}
    S = \int d^4 x \, \sqrt{-\widetilde{g}} \, \frac{M_{\text{Pl}}^2}{2} 
    \Bigg[ \widetilde{\phi}\,\widetilde{R} 
    -\frac{\omega(\widetilde{\phi})}{\widetilde{\phi}} 
    \widetilde{g}^{\mu\nu}\widetilde{\nabla}_{\mu}\widetilde{\phi}
    \widetilde{\nabla}_{\nu}\widetilde{\phi} 
    \\
    - U(\widetilde{\phi}) \Bigg]
    + \int d^4x \,\sqrt{-\widetilde{g}}\, 
    \mathcal{L}_m\!\left(\psi_m,\widetilde{g}_{\mu\nu}\right) \,.
    \label{eq:jf}
\end{multline}
Above, $\widetilde{g}$ is the metric determinant, $\widetilde{R}$ the Ricci scalar, $M_{\text{Pl}}$ the reduced Planck mass, $\omega(\widetilde{\phi})$ and $U(\widetilde{\phi})$ the coupling and potential, respectively, and $\mathcal{L}_m$ the matter Lagrangian. A conformal transformation, $\tilde g_{\mu\nu}=F^2(\phi)g_{\mu\nu}$, together with a field redefinition, brings the theory into the Einstein frame (EF) \cite{Faraoni1999}:
\begin{multline}
    S = \int d^4x\,\sqrt{-g} \left [ \frac{M_{\text{Pl}}^2}{2}\mathcal{R}-\frac{1}{2}g^{\mu\nu}\nabla_{\mu}\phi\nabla_{\nu}\phi -V(\phi) \right ]
    \\
    + \int d^4x \sqrt{-g}\,\mathcal{L}_m \left ( \psi_{m},F^2(\phi)g_{\mu\nu} \right ) \,.
    \label{eq:ef}
\end{multline}
In this representation, the gravitational action has the canonical Einstein–Hilbert form, while matter couples through the conformal factor $F(\phi)$; the EF scalar potential is denoted as $V(\phi)$ and the Ricci scalar as $\mathcal{R}$.

The first step is to solve the equations for the scalar field in the EF in the case of an extended, static, spherically symmetric body of EF mass $M_E$ and radius $R$, embedded in a homogeneous background of density $\rho_\infty$. The scalar field obeys
\begin{equation}
g^{\mu\nu}\nabla_{\mu}\nabla_{\nu}\phi = \frac{dV_{\text{eff}}}{d\phi}\,,
\label{eq:scalareom}
\end{equation}
where the effective potential is $V_{\text{eff}}(\phi) = V(\phi) +\rho F(\phi)$, with $\rho$ being the conserved Einstein frame density. The value of the scalar field at the minimum of the potential is denoted as $\phi_{\min}(\rho)$, while its corresponding effective mass is
\begin{equation}
    m_{\text{eff}}^2 \equiv \left.\frac{d^2V_{\text{eff}}}{d\phi^2}\right|_{\phi_{\min}}\,.
    \label{eq:effectivemass}
\end{equation}
In particular, $\phi_\infty$ denotes the value minimizing $V_{\text{eff}}$ at $\rho=\rho_\infty$, with the associated scalar mass $m_\infty$, while $\phi_c$ the field value at the minimum of $V_{\text{eff}}$ inside the source of density $\rho_c$, with the associated scalar mass $m_c$. Working in isotropic coordinates and denoting the EF radial coordinate as $\xi$, the exterior solution at 1PN order is
\begin{equation}
    \phi(\xi) \simeq \phi_{\infty} - \epsilon\, M_{\text{Pl}}\,\frac{G M_E}{\xi}\,e^{-m_\infty \xi}\,,
\label{eq:sfsol}
\end{equation}
where the scalar charge $\epsilon$, which controls the efficiency of screening, is defined as
\begin{align} 
    \epsilon &\equiv \frac{\phi_{\infty} - \phi_{c}}{M_{\text{Pl}}\,\Phi_{\text{N}}}\,, \label{eq:scalarcharge} \\ \Phi_{\text{N}} &= \frac{G M_E}{R}\,,
    \label{eq:NP}
\end{align}
with $R$ being the source’s radius; Eq.~\eqref{eq:sfsol}, is valid in the screened/long-range regime $m_c R \gg 1$ and $m_\infty R \ll 1$. We work in this regime and verify a posteriori across the scanned parameter space that these conditions are satisfied in our case. In what follows, we consider a compact, screened source (Earth) in a dilute background, with $\rho_c \gg \rho_\infty$, hence the thin-shell regime ($\epsilon \ll 1$). The density $\rho_\infty$ will be identified with the local Galactic ambient density.

Using the scalar-field solution obtained above, one can compute the EF metric at the 1PN order and then transform it back to the JF; denoting by $r$ the radial coordinate in the JF and by $M$ the mass of the gravitating source, the line element takes the form
\begin{equation}
    \begin{aligned}
    \widetilde{ds}^2 &= F(\phi)^2ds^2 \label{eq:ds2pot2} \\
    &= - \left ( 1 - \frac{2\,\mathcal{G}(\epsilon)M}{r}\,+\,\beta(\epsilon)\frac{2\,\mathcal{G}^2(\epsilon)M^2}{r^2} \right ) dt^2 \\
    &\quad \quad \quad + \left ( 1\,+\,\gamma(\epsilon)\frac{2\,\mathcal{G}(\epsilon)M}{r} \right ) \left ( dr^2 + r^2d\Omega^2 \right ) \,,
    \end{aligned}
\end{equation}
where $d\Omega^2 \equiv d\vartheta^2 + \sin^2\vartheta\, d\varphi^2$ is the metric on the unit two-sphere. The quantities $\mathcal{G}(\epsilon)$, $\gamma(\epsilon)$, and $\beta(\epsilon)$ are determined by the expansion of the conformal factor $F(\phi)$ and potential $V(\phi)$ around $\phi_\infty$. Their explicit expressions are given as follows:
\begin{align}
    &\mathcal{G}(\epsilon) = G F_{\infty}^2 \left [ 1 + \left ( \frac{F_1M_{\text{Pl}}}{F_{\infty}} - \frac{V_1}{M_{\text{Pl}}m_{\infty}^2}\right)\epsilon \right ] \,,
    \label{eq:Geps} \\
    &\gamma(\epsilon) = 1 - \left ( \frac{2F_1M_{\text{Pl}}}{F_{\infty}} - \frac{3V_1}{2M_{\text{Pl}}m_{\infty}^2} \right ) \epsilon \nonumber \\
    &\quad \quad \quad + \left ( \frac{2F_1^2M_{\text{Pl}}^2}{F_{\infty}^2} - \frac{7V_1F_1}{2m_{\infty}^2F_{\infty}} + \frac{3V_1^2}{2M_{\text{Pl}}^2m_{\infty}^4} \right ) \epsilon^2\,,
    \label{eq:gammaeps} \\
    &\beta(\epsilon) = 1 + \frac{3V_1}{4M_{\text{Pl}}m_{\infty}^2}\epsilon \nonumber \\
    &+ \left [ \left ( \frac{F_2}{F_{\infty}} - \frac{F_1^2}{2F_{\infty}^2} \right ) M_{\text{Pl}}^2 - \frac{3V_1F_1}{2m_{\infty}^2F_{\infty}} + \frac{3V_1^2}{4M_{\text{Pl}}^2m_{\infty}^4} \right ] \epsilon^2\,,
    \label{eq:betaeps}
\end{align}
where we defined
\begin{align}
    F_\infty &\equiv F(\phi_\infty)\,, &V_\infty &\equiv V(\phi_\infty)\,, \\
    F_1 &\equiv \frac{dF(\phi)}{d\phi}\bigg\rvert_{\phi_{\infty}}\,, &V_1 &\equiv \frac{dV(\phi)}{d\phi}\bigg\rvert_{\phi_{\infty}}\,, \\
    F_2 &\equiv \frac{1}{2}\frac{d^2F(\phi)}{d\phi^2}\bigg\rvert_{\phi_{\infty}}\,, &V_2 &\equiv \frac{1}{2}\frac{d^2V(\phi)}{d\phi^2}\bigg\rvert_{\phi_{\infty}}\,.
\end{align}
It is important to clarify that, in the expressions above, $\Phi_N$ designates the Newtonian potential in the EF. Nevertheless, because of the mapping between masses and radii in the EF and JF (see Eq.~(29) in Ref.~\cite{Feleppa2024}) and since $F_\infty$ deviates from unity only negligibly, the Newtonian potentials defined in the two frames are effectively the same. Accordingly, in what follows we will, to avoid notational clutter, simply use $\Phi_N$ to denote the Newtonian potential.

The formulas discussed above can be now specialized to the three screening mechanisms of interest — chameleon, symmetron, and dilaton — by inserting their characteristic potentials and couplings into the general expressions for the PPN parameters.

For comparison with the screening mechanisms discussed below, in the  unscreened regime of scalar-tensor theories with a very light scalar and a constant conformal coupling $\beta_c$, the JF metric reduces to the standard massless scalar–tensor form with an effective Newton constant and parameter $\gamma$~\cite{Davis2018,Benisty2023a,Brax2025}
\begin{equation}
    \mathcal{G}_{\text{unscr}} = G (1 + 2\beta_c^2) \, , \quad \gamma_{\text{unscr}} = 1 -\frac{4\beta_c^2}{1 + 2\beta_c^2} \, ,
\end{equation}
with $\beta_c$ the dimensionless coupling between the scalar field and matter in the unscreened limit; we will see how much the screened regime deviates from these results.

\noindent \textit{Chameleon screening. }The chameleon mechanism, originally proposed in Refs.~\cite{KhouryWeltman2004PRL,KhouryWeltman2004PRD}, screens fifth forces by increasing the scalar’s effective mass with the ambient density, thus rendering the interaction short range in dense environments. The scalar potential and conformal coupling function are given by \cite{KhouryWeltman2004PRL,KhouryWeltman2004PRD}
\begin{align}
    &V(\phi) = \frac{\Lambda^{4+n}}{\phi^n}\,, \\
    &F(\phi) = \exp \left ( \frac{\beta_m \phi}{M_{\text{Pl}}} \right ) \,,
\end{align}
respectively; the constant $\Lambda$ sets the mass scale, $\beta_m > 0$ is a dimensionless matter-coupling, and $n > 0$. Now, expanding around the background $\phi_\infty$ determines the coefficients $F_\infty, F_1, F_2, V_1, V_2$ (here and below, the quantity $V_\infty$ acts as a cosmological constant, which we assume to be cancelled by a field-independent vacuum energy), which upon substitution in Eqs.~\eqref{eq:Geps}, \eqref{eq:gammaeps} and \eqref{eq:betaeps} of the previous section, as well as using Eq.~\eqref{eq:scalarcharge}, yields
\begin{align}
    \mathcal{G}_{\text{cham}} &\simeq G \left ( 1 + \frac{\beta_m \phi_{\infty}}{M_{\text{Pl}}\Phi_{\text{N}}} \right ) \,, \label{eq:Gepscham2} \\
    \gamma_{\text{cham}} &\simeq 1 - \frac{2\beta_m \phi_{\infty}}{M_{\text{Pl}}\Phi_{\text{N}}}\,,
    \label{eq:gammaepscham2} \\
    \beta_{\text{cham}} &\simeq 1 - \frac{3}{4\Phi_{\text{N}}(n + 1)} \left ( \frac{\phi_{\infty}}{M_{\text{Pl}}} \right ) ^2\,,
\label{eq:betaepscham2}
\end{align}
where the field value that minimizes $V(\phi)$ is given by
\begin{equation}
    \phi_\infty \simeq \left ( \frac{n M_{\text{Pl}}\Lambda^{4 + n}}{\beta_m \rho_\infty} \right ) ^{\frac{1}{n + 1}}\,.
\label{eq:phiinftycham2}
\end{equation}
In order to obtain the above expressions, one assumes that $\beta_m \phi/M_{\text{Pl}} \ll 1$ \cite{Brax2010b,Burrage2018}.

\noindent \textit{Symmetron screening. }Proposed in Refs.~\cite{HinterbichlerKhoury2010,Khoury2011}, the symmetron mechanism screens by tying the field’s vacuum expectation value, and hence its matter coupling, to the ambient density:~broken symmetry and coupling in low density, restored symmetry and suppressed coupling in high density. The symmetron mechanism is characterized by the following scalar field potential and conformal coupling function \cite{HinterbichlerKhoury2010,Khoury2011}:
\begin{align}
    V(\phi) &= -\frac{1}{2}\mu^2\phi^2 + \frac{\lambda}{4}\phi^4\,, \\
    F(\phi) &= 1 + \frac{\phi^2}{2M_{\text{sym}}^2}\,.
    \label{eq:V&Fsymm}
\end{align}
Here $\mu$ and $M_{\text{sym}}$ are mass-dimension one parameters, while $\lambda > 0$ is dimensionless. Using Eq.~\eqref{eq:scalarcharge}, and expanding the above quantities around $\phi_\infty$ to obtain the coefficients $F_\infty, F_1, F_2, V_1, V_2$, lead to the following expressions:
\begin{align}
    \mathcal{G}_{\text{sym}} &\simeq G \left ( 1 + \frac{\phi_{\infty}^2}{M_{\text{sym}}^2\Phi_{\text{N}}} \right ) \,,
    \label{eq:Gepssym2} \\
    \gamma_{\text{sym}} &\simeq 1 -     \frac{2\phi_{\infty}^2}{M_{\text{sym}}^2\Phi_{\text{N}}}\,,
    \label{eq:gammaepssym2} \\
    \beta_{\text{sym}} &\simeq 1 + \frac{1}{2} \left ( \frac{\phi_{\infty}}{M_{\text{sym}}\Phi_{\text{N}}} \right ) ^2\,,
    \label{eq:betaepssym2}
    \\
    \phi_{\infty} &= \frac{\mu}{\sqrt{\lambda}} \sqrt{1 - \frac{\rho_{\infty}}{M_{\text{sym}}^2\mu^2}}\,.
    \label{eq:phiinftysymmetron2}
\end{align}
The above results hold in the regime $\rho_{\infty} < M_{\text{sym}}^2\mu^2$. Moreover, we have assumed $\phi \ll M_{\text{sym}}$ \cite{Khoury2011} and, without loss of generality, also that the scalar field background field  is positive (see the discussion below Eq.~(72) in Ref.~\cite{Feleppa2024}).

\noindent \textit{Dilaton screening. }In dilaton models, the scalar–matter coupling is dynamically driven to zero in high-density environments, thereby screening fifth forces \cite{DamourPolyakov1994,Brax20102}. In this case, the scalar potential and the conformal coupling function are \cite{DamourPolyakov1994,Brax20102}
\begin{align}
    V(\phi) &= V_{0}\exp \left ( -\frac{\lambda_{\text{dil}} \, \phi}{M_{\text{Pl}}} \right )\,,
    \label{eq:vdilaton} \\
    F(\phi) &= 1+{A_2}\frac{\phi^2}{2M_{\text{Pl}}^2}\,,
    \label{eq:fdilaton}
\end{align}
respectively, where $V_0$ is a constant with units of energy density, while $\lambda_{\text{dil}}$ and $A_2$ are two dimensionless constants ($A_2$ controls the strength of the coupling to matter); in the strong coupling regime, $\phi/M_{\text{Pl}} \ll 1$ \cite{DamourPolyakov1994,Brax20102}, and the effective gravitational strength and the PPN parameters $\gamma(\epsilon)$ and $\beta(\epsilon)$ take the form
\begin{align}
    \mathcal{G}_{\text{dil}} &\simeq G \left ( 1 + \frac{A_2 \, \phi_\infty^2}{M_{\text{Pl}}^2\Phi_{\text{N}}} \right ) \,,
    \label{eq:Gepsdil2} \\
    \gamma_{\text{dil}} &\simeq 1 - \frac{2 A_2 \, \phi_\infty^2}{M_{\text{Pl}}^2\Phi_{\text{N}}}\,,
    \label{eq:gammaepsdil2} \\
    \beta_{\text{dil}} &\simeq 1 + \frac{A_2}{2} \left ( \frac{\phi_\infty}{M_{\text{Pl}}\Phi_{\text{N}}} \right ) ^2\,,
    \label{eq:betaepsdil2}
\end{align}
respectively, where the field at the minimum of the potential reads
\begin{equation}
    \phi_\infty \simeq \frac{\lambda_{\text{dil}} \, M_{\text{Pl}} V_0 }{A_2 \rho_\infty} \,.
    \label{eq:phimindil2}
\end{equation}

\noindent \textbf{Solar-System tests. }Here we provide brief descriptions of the three Earth-based experiments used to constrain screened modified gravity — GP-B, LAGEOS-2, and the Sagnac experiment described in the main text. Moreover, we derive the three observables associated with these tests. For later convenience, we rewrite the line element \eqref{eq:ds2pot2} as
\begin{equation}
    ds^2 = -(1 + 2\Phi) dt^2 + (1 - 2\Psi)\left(dr^2 + r^2 d\Omega^2\right)\,,
\label{metric}
\end{equation}
where we have defined the two potentials $ \Phi = \Phi(r,\epsilon)$ and $\Psi = \Psi(r,\epsilon)$ as
\begin{align}
    \Phi &= -\frac{\mathcal{G}(\epsilon) M}{r} +  \beta(\epsilon)\frac{\,\mathcal{G}^2(\epsilon)M^2}{r^2}\,, \label{Phi}
    \\
    \Psi &= -\gamma(\epsilon) \frac{\mathcal{G}(\epsilon) M}{r} \label{Psi}\,,
\end{align}
respectively.

\begin{figure}[!t]
\includegraphics[width=0.9\linewidth]{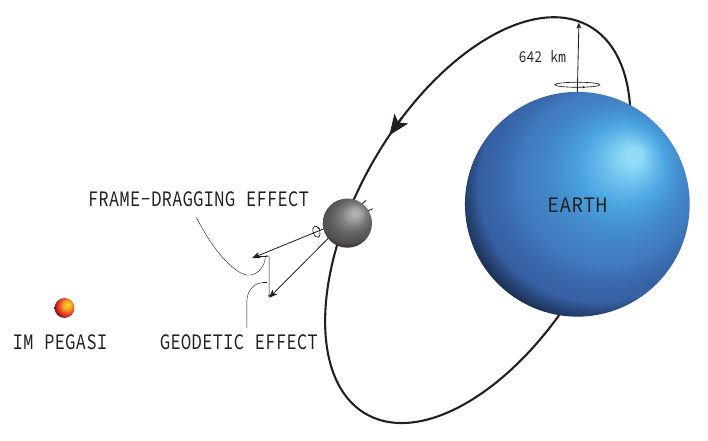}
\centering
\caption{GP-B schematic and geometry of the geodetic and frame-dragging effects. Adapted from Fig.~1 in Ref.~\cite{Everitt2011}.}
\label{fig:GP-B}
\end{figure}
\noindent \textit{GP-B. } The GP-B mission, launched in 2004, was designed to provide a test of Einstein’s General Relativity by directly measuring the geodetic effect and frame-dragging~\cite{Buchman2000,Everitt2015}. The experiment placed four ultra-precise quartz-sphere gyroscopes in a low-Earth polar orbit at an altitude of approximately 640 km; by tracking the orientation of the gyroscopes relative to a distant guide star (IM Pegasi), GP-B measured the relativistic drifts of their spin axes caused by spacetime curvature (geodetic effect) and by Earth’s rotation (frame-dragging), see Fig.~\ref{fig:GP-B} for a schematic. The mission reported a geodetic precession of $-6601.8 \pm 18.3$ milliarcseconds per year and a frame-dragging precession of $-37.2 \pm 7.2$ milliarcseconds per year, both in good agreement with the general relativistic predictions of $-6601$ and $-39.2$ milliarcseconds per year, respectively \cite{Everitt2011}. In what follows, we focus only on the geodetic precession.

The geodetic spin precession of a gyroscope in the spacetime described by the line element \eqref{metric} can be derived as follows.
Consider a test gyroscope on a circular, equatorial geodesic, thus setting
\begin{equation}
    \theta = \frac{\pi}{2}\,, \quad  r = R_{\text{orb}}\,.
\end{equation}
The four-velocity of the gyroscope therefore has only temporal and azimuthal components:
\begin{equation}
    u^\mu = \left(u^0,0,0,u^3\right) = u^0(1,0,0,\Omega)\,, \quad \Omega \equiv \frac{d\phi}{dt}\,.
\end{equation}
Moreover, we introduce its spin four-vector, denoted as \( S^\mu \equiv \big(S^{0}, S^{1}, S^{2}, S^{3}\big) \). In the gyroscope’s instantaneous rest frame, we have $S^\mu = (0,\mathbf{S})$, so that the spin is orthogonal to the four–velocity:
\begin{equation}
    g_{\mu\nu}u^\mu S^\nu = 0\,.
\end{equation}

Working to first order in the potentials, for the line element \eqref{metric} the above constraint gives
\begin{align}
    &-(1 + 2\Phi)\,u^0 S^0 + (1 - 2\Psi) R_{\text{orb}}^2 u^3 S^3 = 0 \nonumber \\
    &\Rightarrow \, \, S^0 \simeq [1 - 2(\Phi + \Psi)] R_{\text{orb}}^2 \, \Omega \, S^3\,,
\label{S0}
\end{align}
with $\Phi = \Phi(R_{\text{orb}},\epsilon)$ and $\Psi = \Psi(R_{\text{orb}},\epsilon)$. The spin vector evolves according to the parallel transport equation
\begin{equation}
    \frac{dS^\mu}{d\tau} + \Gamma_{\alpha \beta}^\mu u^\alpha S^\beta = 0\,.
\end{equation}
At 1PN order, the only relevant non-vanishing Christoffel symbols in our case are 
\begin{align}
    \Gamma_{01}^0 &\simeq \Phi'\,, \\
    \Gamma_{00}^1 &\simeq \Phi'\,, \\
    \Gamma_{33}^1 &\simeq R_{\text{orb}}(R_{\text{orb}}\Psi' - 1)\,, \\
    \Gamma_{31}^3 &\simeq \frac{1}{R_{\text{orb}}} - \Psi'\,,
\end{align}
so that the parallel-transport equations reduce to
\begin{equation}
    \left\{
    \begin{aligned}
    &\frac{dS^0}{d\tau} + \Phi' u^0 S^1 = 0\,,
    \\
    &\frac{dS^1}{d\tau} + \left[R_{\text{orb}}^2 (\Phi' + \Psi') - R_{\text{orb}}\right]\Omega \, u^0 S^3 = 0\,, 
    \\
    &\frac{dS^2}{d\tau} = 0\,,
    \\
    &\frac{dS^3}{d\tau} + \left(\frac{1}{R_{\text{orb}}} - \Psi'\right)\Omega \, u^0 S^1 = 0\,,
    \end{aligned}
    \right.
\label{pte}
\end{equation}
where we used Eq.~\eqref{S0} to eliminate $S^0$ in the evolution equation for $S^1$. Differentiating the second of Eqs.~\eqref{pte} and using the fourth, gives
\begin{equation}
    \frac{d^2 S^1}{dt^2} + \Omega^2 \left(1 - R_{\text{orb}}\Phi' - 2R_{\text{orb}}\Psi'\right)S^1 = 0\,,
\end{equation}
where we also took into account that $u^0 = dt/d\tau$. Setting the initial spin four-vector at $t = 0$ to
\begin{equation}
    S^\mu \rvert_{t = 0} = \left(S^0(0),S^1(0),0,0\right)\,,
\end{equation}
the time evolution equation for $S^1$ can be solved as
\begin{equation}
    \frac{d^2 S^1}{dt^2} + \Omega^2_f \, S^1 = 0 \,\, \Rightarrow \,\, S^1(t) = S^1(0) \cos(\Omega_f t)\,,
\end{equation}
where we defined the quantity
\begin{equation}
    \Omega^2_f \equiv \Omega^2\left[1 - R_{\text{orb}}\left(\Phi' + 2\Psi'\right)\right]\,.
\label{defOmegaf}
\end{equation}
The gyroscope's spin vector precesses in coordinate time at an angular rate $\Omega_f$. In the same time $t = 2\pi/\Omega$, the spin completes an angle
\begin{equation}
    \hat{\alpha} = \Omega_f t = \Omega_f \frac{2\pi}{\Omega}\,.
\end{equation}
Thus, the angle by which the gyroscope's spin axis is displaced after one complete revolution is given by
\begin{equation}
    \alpha = 2\pi - \hat{\alpha} = 2\pi \left(1 - \frac{\Omega_f}{\Omega}\right)\,.
\end{equation}
The ratio $\Omega_f/\Omega$ can be found from Eq.~\eqref{defOmegaf} as
\begin{align}
    \frac{\Omega_f}{\Omega} &= \sqrt{1 - R_{\text{orb}}\Phi' - 2R_{\text{orb}}\Psi'} \nonumber \\
    &\simeq 1 - \frac{1}{2} R_{\text{orb}} \, \Phi' - R_{\text{orb}} \,\Psi' \nonumber \\
    &= 1 - \frac{1 + 2\gamma(\epsilon)}{2}\frac{\mathcal{G}(\epsilon) M}{R_{\text{orb}}}\,,
\end{align}
finally leading to the following expression for the geodetic precession:
\begin{equation}
    \alpha = \pi(1 + 2\gamma(\epsilon))\frac{\mathcal{G}(\epsilon) M}{R_{\text{orb}}}\,,
\label{gprec}
\end{equation}
namely Eq.~(14) in the main text. Note that, to simplify the presentation, the quantity $\epsilon$ is not introduced in the main text.

\noindent \textit{Comparison with previous work. }Here we compare the results obtained in Ref.~\cite{Melville2021} with the ones just obtained. In order to do that, we first specialize our Eq.~\eqref{gprec} to the unscreened regime. In this case, we have
\begin{align}
    &\mathcal{G}(\epsilon) \to \mathcal{G}_{\text{unscr}} = G (1 + 2\beta_c^2)\,, \label{Gunscreened} \\
    &\gamma(\epsilon) \to \mathcal{\gamma}_{\text{unscr}} = \frac{1 - 2\beta_c^2}{1 + 2\beta_c^2}\,, \label{gammaunscreened}
\end{align}
where we recall that $\beta_c$ is the dimensionless coupling between the scalar field and matter in the unscreened limit.
Plugging the above equations into Eq.~\eqref{gprec}, we obtain
\begin{equation}
    \alpha_{\text{unscr}} \simeq 3\pi \frac{G M}{R_{\text{orb}}} - 2\pi \beta_c^2 \frac{G M}{R_{\text{orb}}} \coloneqq \alpha_{\text{GR}} + \Delta \alpha\,.
\label{unscreenedprecession}
\end{equation}
The first term is the pure General Relativity contribution, while the second represents the scalar correction. In Ref.~\cite{Melville2021} , the authors write down the scalar correction to the precession rate per unit time (here denoted as $\Delta \mathbf{\Omega}$) in Eq.~(4.4a). Setting $\Bar{d} = 0$ (no disformal interactions), Eq.~(4.4a) in Ref.~\cite{Melville2021} reduces to
\begin{equation}
    \Delta \mathbf{\Omega} = -\frac{\Bar{c}^2}{M_{\text{Pl}}}\frac{m}{32 \pi r^3}\,,
\end{equation}
which projects to a drift rate (here denoted as $\Delta \dot{s}$) for a circular orbit of radius $r = R_{\text{orb}}$
\begin{equation}
    \Delta \dot{s} = -\frac{\Bar{c}^2}{4}\frac{(G M)^{3/2}}{R_{\text{orb}}^{5/2}}\,.
\end{equation}
Multiplying by the orbital period
\begin{equation}
    T = 2\pi \sqrt{\frac{R_{\text{orb}}^3}{G M}}\,,
\end{equation}
we get the per-orbit angle as
\begin{equation}
    \widetilde{\Delta \alpha} \coloneqq \Delta \dot{s} \, T = -\frac{\pi}{2} \Bar{c}^2 \frac{G M}{R_{\text{orb}}} = -2\pi \beta_c^2 \frac{G M}{R_{\text{orb}}}\,,
\end{equation}
where we also took into account that the quantity $\Bar{c}$ introduced in Ref.~\cite{Melville2021} is related to $\beta_c$ appearing in our Eqs.~\eqref{Gunscreened} and \eqref{gammaunscreened} as $\beta_c = \Bar{c}/2$. The above result matches with the scalar correction obtained in Eq.~\eqref{unscreenedprecession}.

\begin{figure}[!t]
\includegraphics[width=1\linewidth]{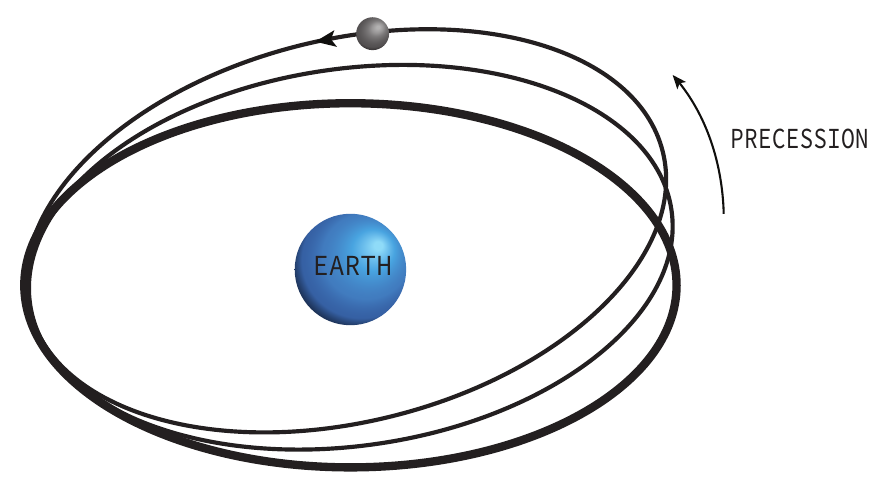}
\centering
\caption{Secular pericenter precession of an Earth-orbiting satellite.}
\label{fig:secularprecession}
\end{figure}
\noindent \textit{LAGEOS-2. }This experiment provides a high-precision orbital of General Relativity by measuring the secular precession of the satellite’s orbital pericenter \cite{Iorio2004,Ciufolini2004,Iorio2009}, see Fig.~\ref{fig:secularprecession} for a schematic of the precession. In principle, the precession receives contributions from three relativistic effects: the Einstein precession (due to Earth’s mass), the de Sitter precession (due to the motion of Earth in the Sun’s field), and the Lense–Thirring precession (caused by Earth’s rotation). However, the Einstein term dominates by a few orders of magnitude \cite{Lucchesi2010}. Consequently, in the following we focus on the Einstein precession, which allows a direct test of the PPN parameters. In order to isolate this effect, in Ref.~\cite{Lucchesi2010}, the authors processed long spans of Satellite Laser Ranging data to LAGEOS-2 with GEODYN II while deliberately omitting relativistic corrections; fitting a linear trend plus a few periodic terms (sinusoids at known thermal/tidal frequencies that average out), they found a secular slope $\Delta\dot{\omega}^{\rm meas} = 3306.58$ milliarcseconds per year, in excellent agreement with the General Relativity expectation $\Delta\dot{\omega}^{\rm GR} \simeq 3305.64$ milliarcseconds per year, with a relative uncertainty of $\sim 2.1\times 10^{-3}$.

The derivation of the secular pericenter–precession rate for a test body orbiting a spherically symmetric mass goes as follows. We start from the metric \eqref{metric} and restrict our attention to equatorial motion by fixing $\theta = \pi/2$. The time-translation and axial Killing symmetries yield the conserved
specific (per-unit-mass) energy and angular momentum
\begin{equation}
     E = (1 + 2\Phi)\,\dot{t}\,, \quad
     L = (1 - 2\Psi) \, r^2 \dot{\phi}\,,
\end{equation}
respectively, where overdots denote derivatives with respect to proper time. The four-velocity normalization 
\begin{equation}
    g_{\mu \nu} u^\mu u^\nu = -1
\end{equation}
yields the radial equation
\begin{equation}
    (1 - 2\Psi)\, \dot{r}^2 = -1 + \frac{E^2}{(1 + 2\Phi)} - \frac{L^2}{(1 - 2\Psi)\,r^2}\,.
\label{radialeqn}
\end{equation}
Making the substitution $u = 1/r$, Eq.~\eqref{radialeqn} becomes
\begin{equation}
   \left(\frac{du}{d\phi}\right)^2 + u^2 = \frac{1 - 2\Psi}{L^2}\left(\frac{E^2}{1 + 2\Phi} - 1\right)\,.
\end{equation}
Differentiating with respect to $\phi$ and expanding consistently at 1PN order gives the perturbed Binet equation
\begin{equation}
    \frac{d^2u}{d\phi^2} + u = \frac{\mathcal{G}(\epsilon) M}{L^2}[1 + (2 - \beta(\epsilon) + 2\gamma(\epsilon))\mathcal{G}(\epsilon) M u]\,.
\end{equation}
Its solution can be written as \cite{Weinberg1972}
\begin{equation}
    \begin{aligned}
    u(\phi) &= \frac{\mathcal{G}(\epsilon) M}{L^2}\left[1 + e\cos((1 - \delta)\phi)\right]\,, \label{u} \\
    L^2 &= \mathcal{G}(\epsilon) M a (1 - e^2)\,, \\
    \delta &\coloneqq \frac{\mathcal{G}(\epsilon) M (2 + 2\gamma(\epsilon) - \beta(\epsilon))}{a (1 - e^2)}\,,
    \end{aligned}
\end{equation}
where $a$ is the semi-major axis of the ellipse and $e$ is the eccentricity of the orbit; the per-orbit advance follows as
\begin{equation}
    \Delta\omega = 2\pi \delta = \frac{6\pi\,\mathcal{G}(\epsilon) M}{a(1 - e^2)}\frac{2  + 2\gamma(\epsilon) - \beta(\epsilon)}{3}\,,
\end{equation}
and converting to a rate using the Newtonian period 
\begin{equation}
    T = \frac{2\pi}{\sqrt{\mathcal{G}(\epsilon) M/a^3}} \, ,
\end{equation}
finally gives 
\begin{equation}
    \Delta \dot{\omega} = \frac{\Delta \omega}{T} = \frac{3\left(\mathcal{G}(\epsilon)M\right)^{3/2}}{a^{5/2}(1-e^2)}\frac{2+2\gamma(\epsilon) - \beta(\epsilon)}{3}\,.
\end{equation}

\begin{figure}[!t]
\includegraphics[width=0.85\linewidth]{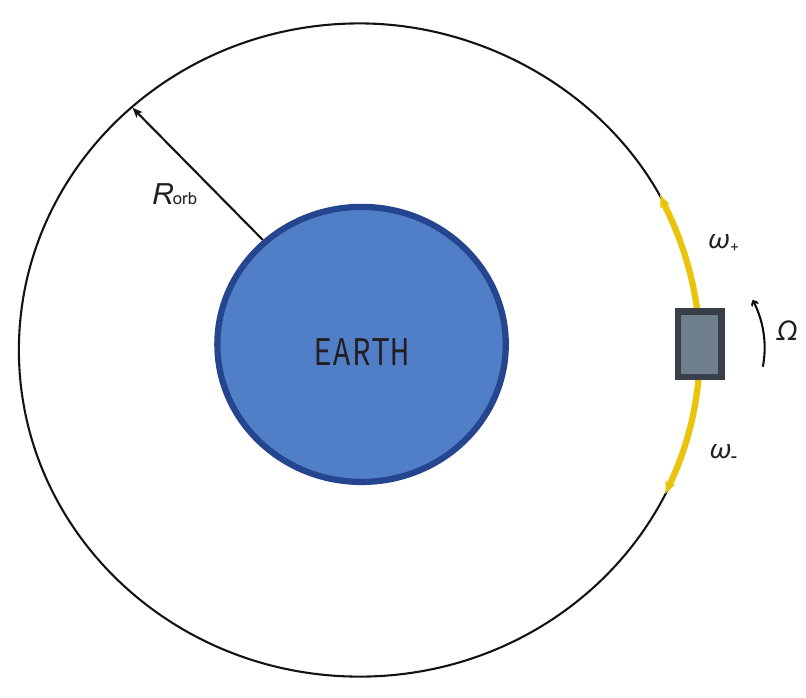}
\centering
\caption{Schematic of a Sagnac experiment around Earth. We denote by $R_{\text{orb}}$ the radius of the circular path, by $\Omega$ the satellite’s angular speed, and by $\omega_{+}$ and $\omega_{-}$ the angular velocities of the co- and counter-rotating light beams. In practice, the ‘ring’ may be approximated by a polygon formed by an orbital relay chain:~satellites on the same orbit forward phase-locked signals around the Earth in both directions, enabling a closed-path measurement referenced to an onboard clock. We also note that reaching an accuracy of $10^{-15}$ for such an experiment requires controlling the resulting path-length difference at the picometer level, so our estimate should be viewed as an idealized sensitivity projection. As a benchmark, mm-class relative position knowledge has been demonstrated in Earth-orbit formation flying (e.g., TanDEM-X \cite{Jäggi2012,Zhang2022} and PROBA-3 \cite{Loreggia2021,Noce2021}), while nm-scale inter-satellite range metrology is available via laser interferometry (for example, GRACE-FO \cite{Abich2019,Ghobadi2022}), which can already be used to monitor and calibrate residual path-length variations.}
\label{fig:Sagnac}
\end{figure}

\noindent \textit{Sagnac time delay. }The Earth-orbiting Sagnac experiment setup was already described in the main text. Here we derive the expression for the Sagnac time delay used to constrain the parameter space of screening models. A visual representation of the setup is shown in Fig.~\ref{fig:Sagnac}.

Let us consider an apparatus rotating with constant angular speed $\Omega > 0$ relative to the static frame, and introduce a co-rotating angular coordinate $\phi' = \phi - \Omega t$. Setting $r=R_{\text{orb}}$ and in the equatorial plane $\theta=\pi/2$, the metric reduces to
\begin{equation}
    ds^2 = g_{00}\,dt^2 + 2g_{03}\,d\phi'\,dt + g_{33}\,d\phi'^2\,,
\end{equation}
with the metric coefficients given by
\begin{align}
    g_{00} &= -\left[1 + 2\Phi - \left(1 - 2\Psi\right) R_{\text{orb}}^2 \, \Omega^2\right]\,, \\
    g_{03} &= \left(1 - 2\Psi\right) R_{\text{orb}}^2 \, \Omega\,,\\
    g_{33} &= \left(1 - 2\Psi\right) R_{\text{orb}}^2\,,
\end{align}
where $\Phi = \Phi(R_{\text{orb}},\epsilon)$ and $\Psi = \Psi(R_{\text{orb}},\epsilon)$.
Light confined on the ring satisfies
\begin{equation}
    ds^2 = 0 \,\, \Rightarrow \,\, g_{00}\,dt^2 + 2g_{03}\,d\phi'\,dt + g_{33}\,d\phi'^2 = 0\,.
\end{equation}
Denoting the photon angular velocity by $\omega \coloneqq d\phi'/dt$, the above equation yields
\begin{equation}
    \omega_{\pm} = \frac{-g_{03} \pm \sqrt{g_{03}^2 - g_{00}g_{33}}}{g_{33}}\,.
\end{equation}
The two roots correspond to the co-rotating ($\omega_+ > 0$) and counter-rotating ($\omega_- < 0$) beams. The coordinate time over one full revolution ($\Delta \phi' = 2\pi$) for each root is
\begin{equation}
    t_{\pm} = \int_0^{2\pi} \frac{d\phi'}{|\omega_{\pm}|}\,.
\end{equation}
Thus, the Sagnac coordinate-time difference is written as
\begin{align}
    \Delta t \coloneqq t_+ - t_- &= \int_0^{2\pi} \left(\frac{1}{\omega_+} + \frac{1}{\omega_-}\right)d\phi' \nonumber
    \\
    &= -4\pi\,\frac{g_{03}}{g_{00}}\,.
\end{align}
For a co-rotating detector we have $d\phi' = 0$, so that we can easily find the Sagnac proper-time difference as
\begin{align}
    \Delta \tau &= \sqrt{-g_{00}} \, \Delta t \nonumber \\
    &= 4\pi\,\frac{R_{\text{orb}}^2 \, \Omega \, (1 - 2\Psi)}{\sqrt{1\,+\,2\Phi - (1 - 2\Psi) R_{\text{orb}}^2 \, \Omega^2}}\,.
\end{align}
At 1PN, and in the slow-rotation limit ($v \equiv R\,\Omega \ll 1$), the above equation simplifies to
\begin{align}
    \Delta \tau &\simeq 4\pi R_{\text{orb}}^2 \, \Omega \, (1 - \Phi - 2\Psi{}+v^2/2) \nonumber \\
    &= 4\pi R_{\text{orb}}^2 \, \Omega \left[1 + \left(1 + 2\gamma(\epsilon) \right)\frac{\mathcal{G}(\epsilon)M}{R_{\text{orb}}}{}+v^2/2 \right]\,,
\label{Deltatau}
\end{align}
which coincides with Eq.~(18) in the main text. For a discussion on the Sagnac effect in General Relativity, see also Refs.~\cite{Benedetto2019,Benedetto2020}. It is important to point out that, although $\Omega$ is in principle $\epsilon$-dependent because screened gravity modifies the satellite's angular velocity, we treat $\Omega$ as an empirically determined orbit input and compute the Sagnac time delay conditional on the observed loop, thereby isolating the light-propagation test from orbit-dynamics modeling.\color{black}

\noindent \textbf{\textit{Modeling oblateness in Sagnac. }}As discussed in the main text, here we model the Earth’s quadrupole $J_2$ and verify that the fractional impact of the $J_2$ uncertainty on the Sagnac delay — normalized to the General Relativity baseline — is below our clock-limited tolerance $10^{-15}$ In what follows, we suppress the explicit $\epsilon$-dependence in the notation, as it is not relevant to the derivation.

To model Earth’s oblateness, we modify the potentials in Eqs. \eqref{Phi} and \eqref{Psi} by the substitution \cite{Ashby2003}
\begin{equation}
    \frac{G M}{r} \,\, \to \,\, \frac{G M }{r}\left[1 - J_2\left(\frac{R}{r}\right)^2 P_2\left(\cos\theta\right)\right]\,,
\label{modifiedpot}
\end{equation}
where $R$ denotes the radius of the oblate body, $J_2$ is a dimensionless measure of its quadrupole moment, and $P_2(\cos\theta)$ is the second-order Legendre polynomial. Plugging the modified potential in Eq.~\eqref{modifiedpot} into Eq.~\eqref{Deltatau}, and setting $r = R_{\text{orb}}$, $\theta = \pi/2$, we obtain
\begin{equation}
    \widetilde{\Delta \tau} = \Delta \tau + \Delta \tau_{J_2}\,,
\end{equation}
where we defined
\begin{equation}
    \Delta \tau_{J_2} \coloneqq 2\pi G M R_{\text{orb}} \, \Omega \, (1 + 2\gamma)\,J_2 \left(\frac{R}{R_{\text{orb}}}\right)^2\,.
\label{DeltatauJ2}
\end{equation}
Let $\Bar{J}_2$ denote the nominal quadrupole coefficient and $\delta J_2$ its uncertainty; we propagate this parameter uncertainty to the observable by a first-order expansion of $\Delta \tau_{J_2}$ around $\Bar{J}_2$, writing
\begin{align}
    \Delta \tau_{J_2} \simeq \Delta \tau_{J_2}\big\rvert_{\Bar{J}_2} + \frac{\partial \Delta \tau_{J_2}}{\partial J_2}\Big\rvert_{\Bar{J}_2}\, \delta J_2 \nonumber \\
    \coloneqq \Delta \tau_{J_2}\big\rvert_{\Bar{J}_2} + \delta \left(\Delta \tau_{J_2}\right)\,.
\end{align}
Using Eq.~\eqref{DeltatauJ2} and $\Delta\tau_{\text{GR}} = \Delta\tau\rvert_{\gamma \to 1}$, we then find
\begin{equation}
    \begin{aligned}
    &\frac{\delta \left(\Delta \tau_{J_2}\right)}{\Delta \tau_{\text{GR}}} \simeq \mathcal{C}\,\frac{\delta J_2}{\Bar{J}_2}\,, \\
    &\mathcal{C} \coloneqq \frac{3}{2}\frac{G M}{R_{\text{orb}}} \Bar{J}_2 \left(\frac{R}{R_{\text{orb}}}\right)^2\,,
    \end{aligned}
\end{equation}
where we normalized on the same fractional scale as our bound, see Eq.~(19). We now require that this fractional impact be below the adopted clock-limited tolerance:
\begin{equation}
    \mathcal{C}\,\frac{\delta J_2}{\Bar{J}_2} \le 10^{-15}\,,
\end{equation}
which, for an orbital radius corresponding to an altitude of $\sim 500\,\text{km}$ above Earth’s surface and $\Bar{J}_2 \approx 10^{-3}$ \cite{Paris2024}, yields $\mathcal{C} \approx 8\times 10^{-13}$, and hence
\begin{equation}
    \frac{\delta J_2}{\Bar{J}_2} \lesssim 10^{-3}\,.
\end{equation}
The quadrupole bound obtained above is comfortably satisfied by the Earth's oblateness value $\Bar{J}_2$ and its uncertainty reported in Ref.~\cite{Paris2024} ($\delta J_2 / \Bar{J}_2 \approx 10^{-7}$); accordingly, we retain a conservative $10^{-15}$ relative uncertainty.

\end{document}